\documentclass[prb,draft,amsfonts,amssymb,amsmath,showpacs]{revtex4}
\begin{document}

\title{Dynamical CP-Violation: A Supplement to the Kobayashi-Maskawa Matrix in the Standard Model}
\author{Tadafumi Ohsaku}
\affiliation{Institut f\"{u}r Theoretische Physik, Universit\"{a}t zu K\"{o}ln, 50937 K\"{o}ln, Deutschland}
\date{\today}

\newcommand{\bmx}{\mbox{\boldmath $x$}}
\newcommand{\bmp}{\mbox{\boldmath $p$}}
\newcommand{\bmk}{\mbox{\boldmath $k$}}
\newcommand{\kfey}{\ooalign{\hfil/\hfil\crcr$k$}}
\newcommand{\pfey}{\ooalign{\hfil/\hfil\crcr$p$}}
\newcommand{\partfey}{\ooalign{\hfil/\hfil\crcr$\partial$}}
\newcommand{\hfey}{\ooalign{\hfil/\hfil\crcr$h$}}

\pacs{11.15.Ff,12.60.-i,12.60.Rc,14.65.-q}

\maketitle

The Kobayashi-Maskawa matrix is a great achievement in the modern particle physics~[1],
and it was discovered for obtaining an explanation of CP-violating processes in phenomenology.
The Kobayashi-Maskawa theory investigated a phase which cannot be absorbed by a redefinition of quark fields described by the framework of the standard model.
The standard model is the most important success in modern particle physics~[2].
In the theories beyond the standard model, the top-condensation model is an interesting candidate for us~[3,4].
The top-condensation model uses a generalized Nambu$-$Jona-Lasinio ( NJL ) model~[5], a four-fermion contact interaction model, 
and thus it is non-renormalizable, though the theory argues that the low-energy effective theory of the top-condensation model 
corresponds to the standard model~[4]. 
Hence the predictions of masses of the model have been examined by employing the renormalization group analysis
which reduces the sensitivity on cutoff and model parameters. 
In this paper, we show another ( can be a supplement ) way with the Kobayashi-Maskawa theory for a CP-violation.
To avoid a possible confusion of readers, we emphasize that we consider a single-flavor model. 
Thus our mechanism of CP-violation takes place without a flavor-violation/oscillation.
( For the conservation of $U(1)$-gauge symmetry in single-flavor quark, 
it is favorable that our mechanism should be generalized into a multi-flavor system. )

\vspace{5mm}

${\bf Assumption}$: Quark fields have very tiny Majorana masses, in the left-right {\it asymmetric} manner.
( Explanation of the "asymmetry" will be given later. )
Usually in particle phenomenology, a super-heavy Majorana mass term is assumed for neutrino for realizing the seesaw mechanism~[6-8].
Here, we consider that the Majorana masses of quarks are very tiny. 
The origin of the Majorana masses are set aside for a while.
It might be given by a VEV of a Higgs field, generated by an interaction of an underlying theory, 
a compactification of strings, some effects of extra dimensions, so forth.

\vspace{3mm}

${\bf Proposition}$: Quarks will have a spontaneous ( dynamical ) CP-violation as a result 
of a dynamical generation of a Dirac mass terms under the NJL mechanism.
In other words, quarks have a mechanism of dynamical chiral mass generation accompanied with a spontaneous CP-violation.

\vspace{3mm}

${\bf Proof}$: Evident from some parts of the results in Ref.~[9].

\vspace{5mm}

Here, we explain the situation considered here by the results in Ref.~[9].
We introduce the following Lagrangian for quarks:
\begin{eqnarray}
{\cal L} &=& {\cal L}_{F} - V(H,H^{\dagger}),   \\
{\cal L}_{F} &=& \xi^{\dagger}i\sigma^{\mu}\partial_{\mu}\xi -\eta^{\dagger}i\sigma^{\mu}\partial_{\mu}\eta
-\frac{1}{2}\bigl( M_{R}\xi^{\dagger}i\sigma_{2}\xi^{*}-M^{\dagger}_{R}\xi^{T}i\sigma_{2}\xi \bigr)
-\frac{1}{2}\bigl( M_{L}\eta^{T}i\sigma_{2}\eta-M^{\dagger}_{L}\eta^{\dagger}i\sigma_{2}\eta^{*} \bigr) +G\eta^{\dagger}\xi\xi^{\dagger}\eta,     \\
M_{R} &=& g_{R}\langle H\rangle, \quad M_{L} = g_{L}\langle H \rangle.
\end{eqnarray}
Here, the Lagrangian is a one-flavor model. $\xi$ and $\eta$ are right- and left- handed Weyl spinors, respectively.
$V(H,H^{\dagger})$ is a Higgs potential, though it is not a necessity for our model that $H$ is a Higgs field in principle ( see the Assumption given above ).
The phases of the mass parameters $M_{R}$ and $M_{L}$ can be absorbed by the redefinition of the phases of $\xi$- and $\eta$- fields when
the theory does not generate a Dirac mass term.
Just the same with the case of the Nambu$-$Jona-Lasinio model, the four-fermion interaction can generate a Dirac mass term as follows:
\begin{eqnarray}
G\eta^{\dagger}\xi\xi^{\dagger}\eta &\to& -m_{D}\xi^{\dagger}\eta-m^{\dagger}_{D}\eta^{\dagger}\xi 
= -m_{D}\bar{\psi}P_{+}\psi - m^{\dagger}_{D}\bar{\psi}P_{-}\psi,    \\
& & P_{R} = \frac{1}{2}(1+\gamma_{5}), \quad P_{L} = \frac{1}{2}(1-\gamma_{5}).
\end{eqnarray}
The final expression is given in terms of Dirac bispinors.
Obviously, this Dirac mass term at $m\ne m^{\dagger}$ breaks the CP-symmetry,
and now the phases of $M_{R}$, $M_{L}$ and $m_{D}$ cannot be absorbed simultaneously,
if $M_{R}\ne M_{L}$ ( more precisely, under the case where the phase of $M_{R}$ coincides with that of $M_{L}$ ).
The fermion sector is written down in the following matrix form:
\begin{eqnarray}
{\cal L}_{F} &=& \frac{1}{2}\overline{\Psi_{MN}}\Omega_{M}\Psi_{MN},   \\
\Omega_{M} &\equiv& \left(
\begin{array}{cc}
i\partfey -M^{\dagger}_{R}P_{+} -M_{R}P_{-} & -m^{\dagger}_{D}P_{+} -m_{D}P_{-}  \\
-m^{\dagger}_{D}P_{+} -m_{D}P_{-}  & i\partfey -M^{\dagger}_{L}P_{+} -M_{L}P_{-} 
\end{array}
\right), \\
\Psi_{MN} &\equiv& \left(
\begin{array}{c}
\psi_{MR} \\
\psi_{ML}
\end{array}
\right), \quad \overline{\Psi_{MN}} = (\overline{\psi_{MR}},\overline{\psi_{ML}}) \quad \psi_{MR} = \left(
\begin{array}{c}
\xi \\
i\sigma_{2}\xi^{*}
\end{array}
\right), \quad \psi_{ML} = \left(
\begin{array}{c}
-i\sigma_{2}\eta^{*} \\
\eta
\end{array}
\right).
\end{eqnarray}
Here, we have used the Majorana-Nambu notation~[9,10] $\Psi_{MN}$ defined by Majorana fields $\psi_{MR}$ and $\psi_{ML}$, 
and of course it is equivalent with.the Dirac field expression ( Dirac-Nambu notation )~[10-14].
By the result of diagonalization of the matrix $\Omega_{M}$, the energy-momentum relation of fermions will be obtained in the following form:
\begin{eqnarray}
E_{\pm}(\bmk) &=& \sqrt{\bmk^{2}+M^{2}_{\pm}}, \\
M_{\pm} &=& \sqrt{|m_{D}|^{2} + \frac{|M_{R}|^{2}+|M_{L}|^{2}}{2}\mp \frac{1}{2}\sqrt{ (|M_{R}|^{2}-|M_{L}|^{2})^{2} + 4|m_{D}|^{2}(|M_{R}|^{2}+|M_{L}|^{2}+2|M_{R}||M_{L}|\cos\Theta) } }, \\
\Theta &\equiv& \theta_{R} + \theta_{L} -2\theta_{D}.
\end{eqnarray}
Here, the definition of phases of the mass parameters are
\begin{eqnarray}
M_{R} = |M_{R}|e^{i\theta_{R}}, \quad M_{L} = |M_{L}|e^{i\theta_{L}}, \quad m_{D} = |m_{D}|e^{i\theta_{D}}.
\end{eqnarray}
The diagonalization will be achieved by several ways ( for example, Ref.~[13] ). 
The energy-momentum relations $E_{\pm}(\bmk)$ have the phase $\Theta$ which cannot be absorbed by a redefinition of fields $\xi$ and $\eta$. 
At first glance, the energy-momentum spectra seems to have the complicated structures, though the spectra do not break the proper Lorentz symmetry $O(3,1)$.
Similar spectra appear in the mass spectra of top-bottom ( stop-sbottom ) in the theory of (minimal supersymmetric) standard model ( for example, Ref.~[15] ),
though the situation we consider here is very different from such a model.
Under our assumption, we argue that the observed quark masses can have the structures given by $M_{\pm}$, constructed by the very-tiny Majorana mass parameters $M_{R}$, $M_{L}$
and the dynamically generated Dirac mass parameter $m_{D}$. 
If we regard $M_{\pm}$ as the observed masses, surely they must be almost degenerate ( quasi-degenerate ).
The actual situation of quark masses would be the superposition of the chiral symmetry breaking of QCD, flavor-violation by the Kobayashi-Maskawa matrix, 
and the effect we consider here ( of course, our model does not give a flavor violation ).
Our model given above can easily be generalized to have both a flavor-degree of freedom and electroweak gauge symmetry, by the similar way with the standard model
( our model can be derived by a decomposition of doublets of the Yukawa term of the standard model ), and the intra- and inter- flavor inteactions
( inter-flavor mechanism seems more favorable due to charge conservation ).
An extension to a supersymmetric model is also possible by employing a similar formalism of supersymmetric BCS superconductivity~[13,14].
Especially, by a more concrete set up of our model, an evaluation of scattering amplitude or a decay constant are intersting for us.
If our story is true, some cosmological observations could prove it.

\end{document}